\documentclass[RNAAS]{aastex62}

\begin{document}

\title{New Jupiter Satellites and Moon-Moon Collisions}

\author[0000-0003-3145-8682]{Scott S. Sheppard}
\affil{Department of Terrestrial Magnetism, Carnegie Institution for Science, 5241 Broad Branch Rd. NW, Washington, DC 20015, USA, ssheppard@carnegiescience.edu}

\author{Gareth V. Williams}
\affil{Minor Planet Center, Harvard-Smithsonian Center for Astrophysics, 60 Garden Street, Cambridge, MA 02138, USA}

\author{David J. Tholen}
\affil{Institute for Astronomy, University of Hawai'i, Honolulu, HI 96822, USA}

\author{Chadwick A. Trujillo}
\affil{Northern Arizona University, Flagstaff, AZ 86011, USA}

\author{Marina Brozovic}
\affil{Jet Propulsion Laboratory, California Institute of Technology, 4800 Oak Grove Drive, Pasadena, CA 91109, USA}

\author{Audrey Thirouin}
\affil{Lowell Observatory, 1400 W Mars Hill Road, Flagstaff, AZ 86001, USA}

\author{Maxime Devogele}
\affil{Lowell Observatory, 1400 W Mars Hill Road, Flagstaff, AZ 86001, USA}

\author{Dora Fohring}
\affil{Institute for Astronomy, University of Hawai'i, Honolulu, HI 96822, USA}

\author{Robert Jacobson}
\affil{Jet Propulsion Laboratory, California Institute of Technology, 4800 Oak Grove Drive, Pasadena, CA 91109, USA}

\author{Nicholas A. Moskovitz}
\affil{Lowell Observatory, 1400 W Mars Hill Road, Flagstaff, AZ 86001, USA}

\keywords{planets and satellites: individual (S/2016 J1, S/2016 J2, S/2017 J1, S/2017 J2, S/2017 J6, S/2018 J1)}

\section{} 

We report the discovery of 12 satellites of Jupiter, giving Jupiter 79
known satellites.  They are between 23rd$-$24th mag in the r-band and
$1-3$ km assuming dark albedos.  Most were discovered using DECam on
the Blanco 4m telescope in March 2017 during a continuation of the
outer solar system survey detailed in \cite{2016AJ....152..221S} and
\cite{2016ApJ...825L..13S}.  Recoveries were obtained at the Magellan,
Discovery Channel, Subaru and Gemini telescopes.  S. Sheppard
maintains a
\href{https://sites.google.com/carnegiescience.edu/sheppard/moons/jupitermoons}{webpage
  showing the characteristics of the moons}.

\begin{figure}
\vspace*{0.5in}
\centerline{\includegraphics[angle=90,totalheight=0.5\textheight]{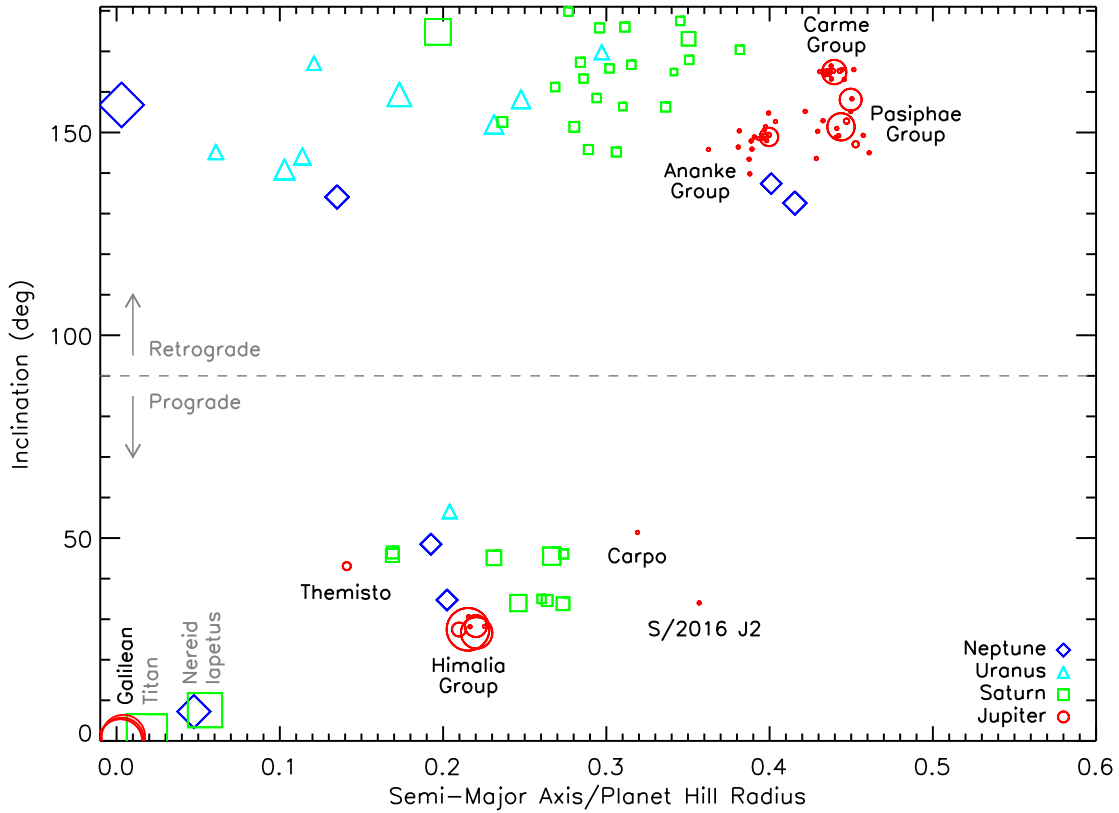}}
\caption{The outer satellites with well determined orbits and large
  inner satellites. Symbol size represents the Log of the satellite size.}
\label{fig:Jupiterai}
\end{figure}

Nine of the discoveries are in the distant retrograde groupings (Fig
\ref{fig:Jupiterai}).  The retrogrades are clustered into orbital
groupings that might be the remnants of once-larger parent bodies that
fragmented from collisions with asteroids, comets, or other satellites
(\cite{2003Natur.423..261S}, \cite{2004AJ....127.1768N},
\cite{2007ARA&A..45..261J}, \cite{2008ssbn.book..411N},
\cite{2010AJ....139..994B}, \cite{2014ApJ...784...22N}).  Three are in
the compact Carme group (S/2017 J5, S/2017 J8 and S/2017 J2).  Four
are in the more diffuse Ananke group (S/2017 J7, S/2016 J1, S/2017 J3
and S/2017 J9), where Euporie is somewhat disconnected.  Two are in
the Pasiphae group (S/2017 J6 and S/2017 J1), which is the most
dispersed grouping, with S/2017 J6 having the largest eccentricity,
0.557, and aphelion at 0.66 Hill radii.  Jupiter LVIII (S/2003 J15) is
somewhat disconnected from the group with a lower inclination and
semi-major axis.

Two are in the closer Himalia prograde group near 28$^\circ$
inclination (S/2017 J4 and S/2018 J1).  These were the brightest found
around 23rd mag, but were harder to discover and track because they
are more in the glare of Jupiter.  The Himalia group has a large
velocity dispersion \citep{2018Icar..310...77L}, which we suggest
could be explained by a second breakup event after an initial Himalia
breakup as most Himalia members don't cluster near Himalia but further
away, near Elara and Lysithea.

S/2016 J2, nicknamed Valetudo, has an orbit unlike any other and is
the most distant prograde satellite around any planet at 0.36 Hill
radii \citep{2018MPEC....O...09S}.  Numerical simulations show it
stable, with average and range of $i=34.2\pm 3^\circ$, $e=0.216\pm
0.125$, and $a=1.89\pm0.07 \times 10^{7}$ km over $10^{8}$ yrs.  Our
stability simulations show a S/2016 J2 like orbit would be stable out
to $a=2.18 \times 10^{7}$ km or $0.41$ Hill radii, but no further,
unlike more distant and eccentric retrograde satellites.  S/2016 J2's
large semi-major axis means it significantly overlaps the orbits of
the distant retrogrades, unlike most inner progrades.  Carpo also has
significant overlap with the retrogrades, though at a higher
inclination, 51$^\circ$.4, than S/2016 J2 at 34$^\circ$.0.

The retrogrades are not expected to collide much among themselves
\citep{2003AJ....126..398N}.  Using the MERCURY integrator and a
particle in a box calculation, we find S/2016 J2 has at best a few
percent chance of colliding with a big retrograde (Pasiphae, Ananke,
Carme or Sinope) over 4.5 Gyrs.  At 1 km, S/2016 J2 would not disrupt
the bigger retrogrades, though a head-on impact would be very
energetic at several km/s, likely producing several large fragments
\citep{1999Icar..142....5B}.  If S/2016 J2 was several times larger in
the past, it would be more likely to collide with and fragment a large
retrograde satellite.  Though not favorable, its possible S/2016 J2 is
the biggest remnant of a once-larger satellite that helped form a
retrograde cluster through collisions.  Prograde Carpo has similar
retrograde collision probabilities as S/2016 J2, it being bigger but
closer in.  Retrograde collisions with Himalia members are also
possible \citep{2003AJ....126..398N}.  Though retrograde-prograde
moon-moon collisions are unlikely individually, taken together, a
retrograde-prograde collision has likely occurred between the outer
satellites of Jupiter.

We observed every well known outer satellite of Jupiter in 2017 and
2018 and rediscovered many of the 2003 and 2011 satellites that had
poorly determined orbits (\cite{2012AJ....144..132J},
\cite{2017AJ....153..147B}).  We also found many unknown satellites
over 24th mag that were too faint to follow and obtain good
orbits.  Two of these lost satellites appear to be in the Himalia group
and the rest retrogrades.  Some are likely the faint low quality orbit
2003 satellites yet to be linked.  This confirms
\cite{2003Natur.423..261S} that Jupiter has about a hundred satellites
larger than 1 km.

The smallest satellites in the orbital groups are still abundant and
not dispersed towards Jupiter.  This suggests the collisions that
created them occurred after the era of planet formation as significant
gas and dust would preferentially drag the smaller satellites inward
\citep{2005SSRv..116..441J}.

There are no obvious retrograde satellite groups around the other
giant planets, except possibly Neptune \citep{2006AJ....132..171S}.
Saturn's progrades show inclination clustering but with less stringent
semi-major axis clustering (\cite{2001Natur.412..163G},
\cite{2018ApJ...859...97H}).

\acknowledgments

This project used DECam, which was constructed by the Dark Energy
Survey collaboration.  Observations were partly obtained at NOAO and
Gemini that are operated by AURA, under cooperation with NSF.  This
work was partly funded by NASA planetary grant NN15AF44G.

\end{document}